\def\vf{v_F}

\newcommand{\DD}{\Delta_0}
\documentclass{LT23auth}
\usepackage{graphicx}

\begin{document}

\begin{frontmatter}

\title{Tail states in superconductors with weak magnetic impurities}

\author[address1]{I. Vekhter \thanksref{thank1}},
\author[address2]{A. V. Shytov},
\author[address3]{I. A. Gruzberg},
\author[address1]{A. V. Balatsky}

\address[address1]{Theoretical Division, MS B262, 
Los Alamos National Laboratory, Los Alamos, NM 87545}

\address[address2]{Kavli Institute for Theoretical Physics, 
University of California, Santa Barbara, CA 93106
and \\ L.~D.~Landau Institute for Theoretical Physics, 2 Kosygin St.,
Moscow, Russia 117334 }

\address[address3]{Department of Physics, Massachusetts Institute
of Technology, Cambridge, MA 02139}

\thanks[thank1]{Corresponding author. E-mail: vekhter@lanl.gov}

\begin{abstract}
We analyse the behavior of the density of states in a singlet
$s$-wave superconductor with weak magnetic impurities in the clean
limit by using the method of optimal fluctuation. We show that the
density of states varies as $\ln N(E)\propto -|E-\DD|^{(7-d)/4}$
near the mean field gap edge $\DD$ in a $d$-dimensional
superconductor. The optimal fluctuation in $d>1$ is strongly
anisotropic. We compare the density of states with that obtained
in other recent approaches.
\end{abstract}

%
%
\end{frontmatter}


%
%


%
%

Studies of spectral properties of disordered superconductors
remain an active area of research as they help advance our
understanding of the competition between disorder and
interactions. Recent years witnessed a renewed interest in the
behavior of the density of states (DOS) in singlet $s$-wave
superconductors with magnetic impurities. Such impurities are
pairbreaking, and are characterized, in the weak scattering limit,
by the spin flip scattering time, $\tau_s$. In the self-consistent
Born approximation (SCB) \cite{AG}, the dimensionless parameter
controlling the suppression of the single particle spectral gap,
$\DD$, is $\Delta\tau_s$, where $\Delta$ is the amplitude of the
superconducting order parameter. In this paper we consider the
clean limit, $\Delta\tau_s\gg 1$, where the SCB approach yields a
finite spectral gap, $\DD\approx\Delta$, with the DOS $N(E)=0$ at
energies $E<\DD$.

It was argued in Ref.\cite{BT} that rare regions where local
impurity concentration is high enough to locally destroy
superconductivity lead to a finite density of states at the Fermi
level. The argument was similar to the method of optimal
fluctuation (OF), well known from studies of doped semiconductors
\cite{Lifshits}: in averaging over {\it all} the realizations of
the impurity distribution, the probability of finding the
realization which locally destroys the gap determines the residual
DOS. Later, Lamacraft and Simons \cite{Simons} considered in
detail the energy dependence of the DOS below the mean field gap
in a dirty superconductor, where the scattering rate due to
potential scattering greatly exceeds the spin-flip pairbreaking
scattering rate, $1/\tau_s$.

Very recently we analysed the subgap DOS in a clean $s$-wave
superconductor, where the spin-flip scattering is dominant
\cite{Shytov}; we argued that at least in some cases this limit is
relevant experimentally. Here we briefly review the results of
Ref.\cite{Shytov} and then present a more detailed comparison of
the DOS obtained in Refs.\cite{BT,Simons,Shytov}.

We consider a mean field hamiltonian
\begin{equation}
\widehat H=\widehat\xi\tau_3+\Delta({\mathbf{r}})\tau_1\sigma_2
+\widehat U,
\end{equation}
where $\widehat\xi=-\nabla^2/(2m)-\mu$, $\mu$ is the chemical
potential, $\tau_i$ and $\sigma_i$ are the Pauli matrices in the
particle-hole and the spin space respectively. The potential due
to magnetic impurities $\widehat U= {\bf U}({\mathbf{r}})\cdot
\mathbf{s}$, where $\mathbf{s}$ is the electron spin operator,
${\bf U}({\mathbf{r}})=\sum_i J {\bf S}_i\delta({\mathbf{r}} -
\mathbf{r}_i)$, $J$ is the exchange constant, and ${\bf S}_i$ is
the impurity spin at a site $i$.

For an energy, $E<\DD$,  OF is the most probable configuration of
impurities that creates a state at $E$, and therefore contributes
the most to the DOS \cite{Lifshits}. OF provides nonperturbative
corrections to the DOS determined in the framework of SCB. In
essentially all the energy range below the gap the size of the OF
is significantly greater than the distance between impurities, so
that the exact impurity potential can be replaced by a smooth
function, and its probability density is well approximated by a
Gaussian \cite{Lifshits} with a width $U_0^2=n_{imp} J^2
S(S+1)/3$, so that $\tau_s^{-1}=2\pi N_0 U_0^2$, where $n_{imp}$
is the impurity concentration, and $N_0$ is the normal state DOS.
The DOS is then given by $\ln\left[ N(E)/N_0 \right]\approx -
{\textrm S}[{\mathbf U}_{opt}]$, where ${\textrm S}[{\mathbf
U}_{opt}]$ is obtained by minimizing
\begin{equation}
\label{action}
    {\textrm S}[{\mathbf U}]=\frac{1}{2U_0^2}\int d^d{\mathbf r}
{\mathbf U}^2(\mathbf{r}) +
    \lambda \biggl({\textrm E }[{\mathbf U}] -E\biggr)
\end{equation}
with respect to both the potential ${\mathbf U}$ and the Lagrange
multiplier $\lambda$. We assume OF to be ferromagnetic
\cite{Shytov}, so that ${\textrm E}[{\mathbf U}]=\langle\Psi |
\widehat\xi\tau_3+\Delta_0\tau_1 + U |\Psi \rangle$ where $\Psi$
is the normalized spinor wave function of the particle in the OF,
and $U(x)=-\lambda U_0^2 (\Psi^\star(x)\Psi (x))$ is now scalar.

In $d=1$ we  find ($\epsilon=|E|/\DD$)
\begin{eqnarray}\label{OptimalU}
&&\frac{U(x)}{2\DD}=-\frac{1-\epsilon^2}{\epsilon + \cosh
(2x\sqrt{1-\epsilon^2}/\xi_0)},
\\
\label{OptimalAction1D} &&{\mathrm S_0}\equiv{\textrm
S}[U_{opt}]=8\pi (\DD\tau_s) \left[
\sqrt{1-\epsilon^2}-\epsilon\arccos\epsilon\right],
\end{eqnarray}
where $\xi_0=v_F/\DD$ is the coherence length and $v_F$ is the
Fermi velocity.  The size of the OF, $L_x\simeq
\xi_0/\sqrt{1-\epsilon^2}\geq\xi_0$, and the depth of the
potential $|U|\leq 2\DD$. In particular, near the gap edge,
$1-\epsilon \ll 1$, when $|U|\ll\DD$ and $L_x\gg\xi_0$, we find
${\mathrm S_0}\simeq (8\pi/3) (\DD\tau_s) (1-\epsilon^2)^{3/2}$,
which allows for a simple interpretation. In an OF of depth $U$
and size $L_x$ the energy of the bound state is  $E\simeq
U+\DD+v_F^2/L_x^2\DD$, subject to optimization of
Eq.(\ref{action}). This means $|E-\DD|\sim v_F^2/L_x^2\DD\sim
|U|$, from which an estimate for ${\textrm S_0}\simeq
L_xU^2/U_0^2\sim (\DD\tau_s) (1-\epsilon^2)^{3/2}$ follows
immediately.

In $d>1$ and for $1-\epsilon \ll 1$ we must compare the action for
the isotropic OF with $L\sim\xi_0/\sqrt{1-\epsilon^2}$ (when
kinetic energy is $\vf/L$), and for an anisotropic OF with
$L_x\sim L$, and transverse size $L_t\sim(L_x/k_F)^{1/2}$
(corresponding to a wave function $\Psi({\mathbf r})=\exp(i k_F
x)\Phi(x,{\mathbf y})$, where $\Phi$ is a slowly varying function,
so that $1/(mL_t^2)\sim v_F/L_x$). The anisotropic OF is favored
by a factor $(E_F/\DD)^{(d-1)/2} (1-\epsilon)^{-(d-1)/4}$ and
gives \cite{Shytov}
\begin{equation}
    \label{AnisAction}
    {\mathrm S_0}\simeq L_x L_t^{d-1}\frac{U^2}{U_0^2}\simeq
    (\DD\tau_s)\left(\frac{E_F}{\DD}\right)^{\frac{d-1}{2}}
    \left(1-\epsilon\right)^{\frac{7-d}{4}}.
\end{equation}

We now compare this result with that of Ref.\cite{Simons} for the
diffusive propagation of the states within OF. It is clear that
even in the clean case considered here, with dilute magnetic
impurities, a transition to the diffusive regime occurs when the
size of OF $L\geq v_F\tau_s$, or $1-\epsilon\leq
(\DD\tau_s)^{-2}$. In our notations the result of
Ref.\cite{Simons} for $\DD\tau_s\gg 1$ reads ${\mathrm
S_D}=(\DD\tau_s)^{5/3}(E_F/\DD)^{d-1}(1-\epsilon)^{(6-d)/4}$.
Consequently, at the crossover point the action from Eq.
(\ref{AnisAction}) is smaller, ${\mathrm S_D/S_0}\simeq
(E_F/\DD)^{(d-1)/2}(\DD\tau_s)^{7/6}\gg 1$, and the OF found here
corresponds to a greater DOS. As the size of the OF increases even
further, the anisotropic fluctuation becomes insupportable due to
diffusive motion. Nonetheless, even for an isotropic fluctuation,
we find ${\mathrm S_D/S_{iso}}\simeq
(\DD\tau_s)^{2/3}(1-\epsilon)^{(d-2)/4}$, which, at the crossover,
gives $(\DD\tau_s)^{(10-3d)/6}\gg 1$ for $d\leq 3$. Consequently,
the OF discussed here still yields a higher DOS than that for
purely diffusive motion. Therefore we expect that the structure of
the OF near the crossover between the ballistic and diffusive
regimes still resembles closely that given by us above, and that
our results remain at least qualitatively valid there.

Balatsky and Trugman \cite{BT} considered only the DOS at $E=0$,
where they needed a large volume fluctuation, $V\geq \xi^d$, which
is less probable and yields lower DOS than that of Eq.
(\ref{AnisAction}). However, in the same spirit we should
investigate whether local suppression of the gap from $\DD$ to $E$
due to a large number of impurities with {\it uncorrelated} spins
(as opposed to a ferromagnetic OF above) is advantageous. For
$1-\epsilon\ll 1$ the local pairbreaking rate, $\gamma$, needed to
reduce the gap to $E$ is $\gamma\tau_s\approx 1+
(1-\epsilon)(\DD\tau_s)^{2/3}$, and the volume of the region has
to be at least equal to that of the anisotropic OF to avoid high
kinetic energy cost (this is an underestimate since it ignores
proximity coupling to bulk). In that case we obtain the optimal
action ${\mathrm S_{BT}}/{\mathrm S_0} \approx (\DD\tau_s)^{1/3}
(E_F/\DD) \bar c$, where $\bar c=n_{imp}\lambda_F^d$ is the atomic
concentration of impurity atoms. As a result, for realistic values
of $\bar c$ and clean samples ${\mathrm S_{BT}}\gg {\mathrm S_0}$,
and the DOS given by the action in Eq.(\ref{AnisAction}) is
higher.

We conclude that the action obtained in our approach gives the
most optimal fluctuation and the highest DOS compared to other
analyses.

\begin{ack}
This research was supported by the DOE under contract
W-7405-ENG-36, by the NSF Grant PHY-94-07194, and by Pappalardo
Fellowship. We are grateful to the ITP Santa Barbara and Aspen
Center for Physics for hospitality and support.

\end{ack}

%
%

\end{document}